\documentclass[aps,pra,reprint,superscriptaddress,amsmath,amssymb,nofootinbib]{revtex4-2}

\usepackage{amsmath,amssymb,amsthm,bm}
\usepackage{graphicx}
\usepackage{epstopdf}
\usepackage{braket}
\usepackage{xcolor}
\definecolor{ringa}{RGB}{216,40,0}
\definecolor{ringb}{RGB}{0,200,80}
\definecolor{ringc}{RGB}{24,112,208}
\definecolor{ringd}{RGB}{248,192,0}
\newcommand{\ra}{\textcolor{ringa}{a}}
\newcommand{\rb}{\textcolor{ringb}{b}}
\newcommand{\rc}{\textcolor{ringc}{c}}
\newcommand{\rd}{\textcolor{ringd}{d}}

\newtheorem{theorem}{Theorem}
\newtheorem{corollary}{Corollary}
\newcommand{\PT}[1]{^{\mathrm{T}_{#1}}}

\begin{document}

\title{Genuinely multipartite entangled qudits: one rule to link them all}

\author{Gon\c{c}alo M. Quinta}
\email{goncalo.quinta@tecnico.ulisboa.pt}
\affiliation{Quantum Physics of Information Group, Instituto de Telecomunica\c{c}\~{o}es, Lisboa, Portugal}
\author{Rui Andr\'{e}}
\affiliation{Quantum Physics of Information Group, Instituto de Telecomunica\c{c}\~{o}es, Lisboa, Portugal}

\date{\today}

\begin{abstract}
Links of rings offer an intuitive picture of multipartite entanglement. Cutting one ring can free some of the others, just as losing a particle, i.e., tracing it out, can disentangle some of the remaining ones. This analogy classifies multipartite entanglement according to which groups of particles stay entangled when others are lost, a question at the heart of robust quantum networks. However, pure states realizing a given link class were known only in special cases. Here, we close this gap. Recasting link polynomials as hypergraphs, we give an explicit formula for pure qudit states and prove that, for every link class, their entanglement under any pattern of particle loss matches the link exactly. The same construction also controls genuine multipartite entanglement, determining exactly which of the groups that stay entangled are genuinely multipartite entangled, and all of them can be made so. As a consequence, we prove the conjecture that $m$-resistant pure qudit states, which stay entangled after the loss of any $m$ particles but become fully separable after the loss of $m+1$, exist for any number of particles and any $m$.
\end{abstract}

\maketitle

\section{Introduction}

Multipartite entanglement is the resource behind many quantum technologies, from quantum networks and distributed protocols to error correction, and understanding how it is organized is therefore of both fundamental and practical interest. A complete classification, however, is only available for very small systems. Under stochastic local operations and classical communication (SLOCC), three qubits can be entangled in two inequivalent ways \cite{Dur2000} while four qubits require nine families, some of them depending on continuous parameters \cite{Verstraete2002}. Beyond that, the number of classes grows so quickly that an exhaustive classification is neither feasible nor particularly useful. This has motivated coarse-grained classifications, which retain only the features that matter for a given task. Examples include a classification of multipartite entanglement in systems of arbitrary local dimension based on orthogonal arrays~\cite{Seveso2018}, and a study of the role of symmetry in the classification and quantification of multipartite entangled states~\cite{Burchardt2021}. Topological ideas have also entered the picture, from topological quantum field theory \cite{Melnikov2023} to persistent homology \cite{Mengoni2020}.

A particularly natural coarse-graining asks which groups of particles remain entangled when the others are lost. This question is operationally important. In a quantum network, particles can be lost, and parties can drop out or simply decline to act. Once that happens, the remaining parties can only exploit the entanglement present in their reduced state. Whether a given subset of parties can still run an entanglement-based protocol without outside help is therefore decided entirely by the entanglement of that subset after the others are traced out \cite{Quinta2018}. Topological links provide an intuitive language for this question. The similarity between entangled particles and linked rings was first noticed by Aravind \cite{Aravind1997}, who associated the GHZ state with the Borromean rings: measuring one particle disentangles the others, just as cutting one ring frees the rest. Shortly afterwards, van Dam \cite{vanDam2002} described the rings of a link that cannot be pulled apart by families of subsets closed under unions of overlapping members. He argued that every such family can be realized both by a link and by a mixed multiqubit state with distillable entanglement on exactly those subsets, and gave pure three-qubit states for all three-ring patterns. Since the outcome of a measurement depends on the basis, Sugita \cite{Sugita2006} proposed instead to associate the cut of a ring with the basis-independent operation of tracing out the corresponding particle. In \cite{Quinta2018}, this analogy was formalized into a classification scheme. Each particle is a ring, tracing out a particle cuts its ring, and particles that remain entangled correspond to rings that remain linked. The details of how the rings are intertwined are ignored: two links are equivalent if every possible combination of cuts leaves the same rings linked. Related connections between knots, braids, and entanglement have been explored from several other directions, including topological entanglement \cite{Kauffman2002}, braiding quantum gates and solutions of the Yang--Baxter equation \cite{Padmanabhan2020a,Padmanabhan2020b,Padmanabhan2020c,Padmanabhan2021,Zhang2024,Padmanabhan2026}, and knot cabling constructions for GHZ and W states \cite{Bhattacharyya2026}.

Within the link picture, the concept of $m$-resistant states was introduced in \cite{Quinta2019}. Such a state remains entangled after the loss of any $m$ particles, but becomes fully separable once more than $m$ particles are lost. Designing such states was shown to be analogous to designing a link of $N$ rings that stays connected after any $m$ cuts but falls apart after $m+1$, and a constructive solution to the latter problem yielded explicit $N$-particle states with the desired resistance to particle loss \cite{Quinta2019}. Entanglement that survives a controlled amount of loss is precisely what a robust quantum network needs, and the problem has since attracted further constructions based on Dicke states and error-correcting codes \cite{Zhang2025}, as well as graph states \cite{Han2026}. As a by-product of the construction presented here, we show that $m$-resistant pure qudit states exist for every $N$ and $m$, with the remaining particles genuinely multipartite entangled.

These results, however, target specific families of links. A related line of work considered pairwise entanglement only. Plesch and Bu\v{z}ek showed that any graph of entangled pairs, i.e., any prescription of which two-party reduced states are entangled, can be realized by a pure multiqubit state~\cite{Plesch2003}, and weighted versions of such entangled graphs have been used to classify four-qubit entanglement~\cite{Gharahi2016}. The link picture, however, requires the analogous statement for subsets of every size, including patterns such as the Borromean rings, where no pair is entangled but the triple is. The general question therefore remains open: does every link class admit a pure quantum state with exactly its pattern of entanglement under particle loss? Mixed-state representatives of every class were sketched in \cite{vanDam2002}, with weights that must decrease fast enough with the size of each entangled subset. Explicit ones were given in \cite{Quinta2018} for four qubits, with coefficients tuned case by case. Pure states, however, were found only for three qubits \cite{vanDam2002, Quinta2018}. A general pure-state construction matters for two reasons: it shows that every pattern allowed by the link formalism is physically realizable, and it gives an explicit state to prepare whenever a network with a prescribed pattern of shared entanglement is needed.

In this work, we give such a construction for every link class, directly from its polynomial. In Sec.~\ref{sec:poly} we review the polynomial formalism, and in Sec.~\ref{sec:hyper} we recast it in the language of hypergraphs, which identifies the class of a link with the family of ring subsets that stay linked. Section~\ref{sec:construction} presents the construction of the corresponding pure qudit states, illustrated with a four-ring example, and states our main result. Section~\ref{sec:gme} shows that the same construction controls genuine multipartite entanglement, which yields states where every linked subset is genuinely multipartite entangled and settles the existence of $m$-resistant states. Section~\ref{sec:conclusions} concludes, and all proofs are given in Appendix~\ref{app:proof}.

\section{From links to polynomials}\label{sec:poly}

\begin{figure}[h!]
\centering
\includegraphics[width=0.45\columnwidth]{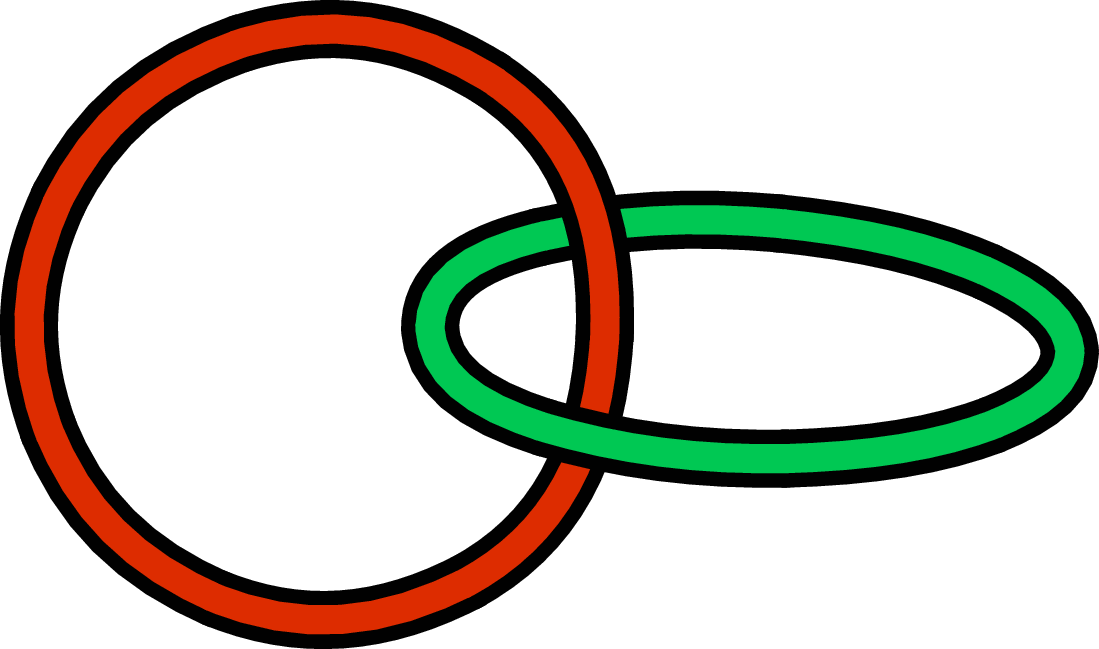}
\caption{The Hopf link, the only element of the class $2^1$, with polynomial $\mathcal{P}(2^1)=\ra\rb$. Ring $\ra$ is drawn in red and ring $\rb$ in green.}
\label{fig:hopf}
\end{figure}

We start by briefly reviewing the polynomial formalism of Ref. \cite{Quinta2018}. The simplest case is two linked rings, the Hopf link\footnote{All link images presented in this work have been created using the \textit{KnotPlot} software. The creation process using the link polynomials is described in \cite{Quinta2018}.} of Fig.~\ref{fig:hopf}, which forms the class $2^1$. To encode it, each ring is assigned a variable, here $a$ and $b$, and the product of variables indicates that the corresponding rings are linked, so $\mathcal{P}(2^1)=ab$. Cutting a ring corresponds to setting its variable to zero. Cutting either ring of the Hopf link gives $\mathcal{P}=0$, meaning that no rings remain linked.

For more rings, a polynomial is a sum of monomials, each representing a set of rings linked together. For $N$ rings there are $2^N-N-1$ possible monomials of order two or higher, but not every sum of them defines a distinct link class. The polynomials are restricted by five rules \cite{Quinta2018}:
\begin{enumerate}
\item No variable appears with a power greater than one. A term such as $aab$ carries no more information than $ab$.
\item Every ring variable appears, and the monomials link all rings together: the variables cannot be split into two groups with no monomial containing variables of both. For example, $ab+cd$, which describes two separate Hopf links, is excluded.
\item There are no first-order terms, since a single ring is not linked to anything.
\item Relabeling the variables does not change the class. For example, $ab+ac$ and $ab+bc$ describe the same link.
\item An $n$-variable monomial $M$ is dropped if its variables are already linked by lower-order monomials built only from the variables of $M$.
\end{enumerate}

\begin{figure}[h]
\centering
\newcommand{\linkfig}[1]{\parbox[c][2.9cm][c]{0.46\columnwidth}{\centering\includegraphics[width=0.42\columnwidth,height=2.8cm,keepaspectratio]{#1}}}
\begin{tabular}{@{}cc@{}}
\linkfig{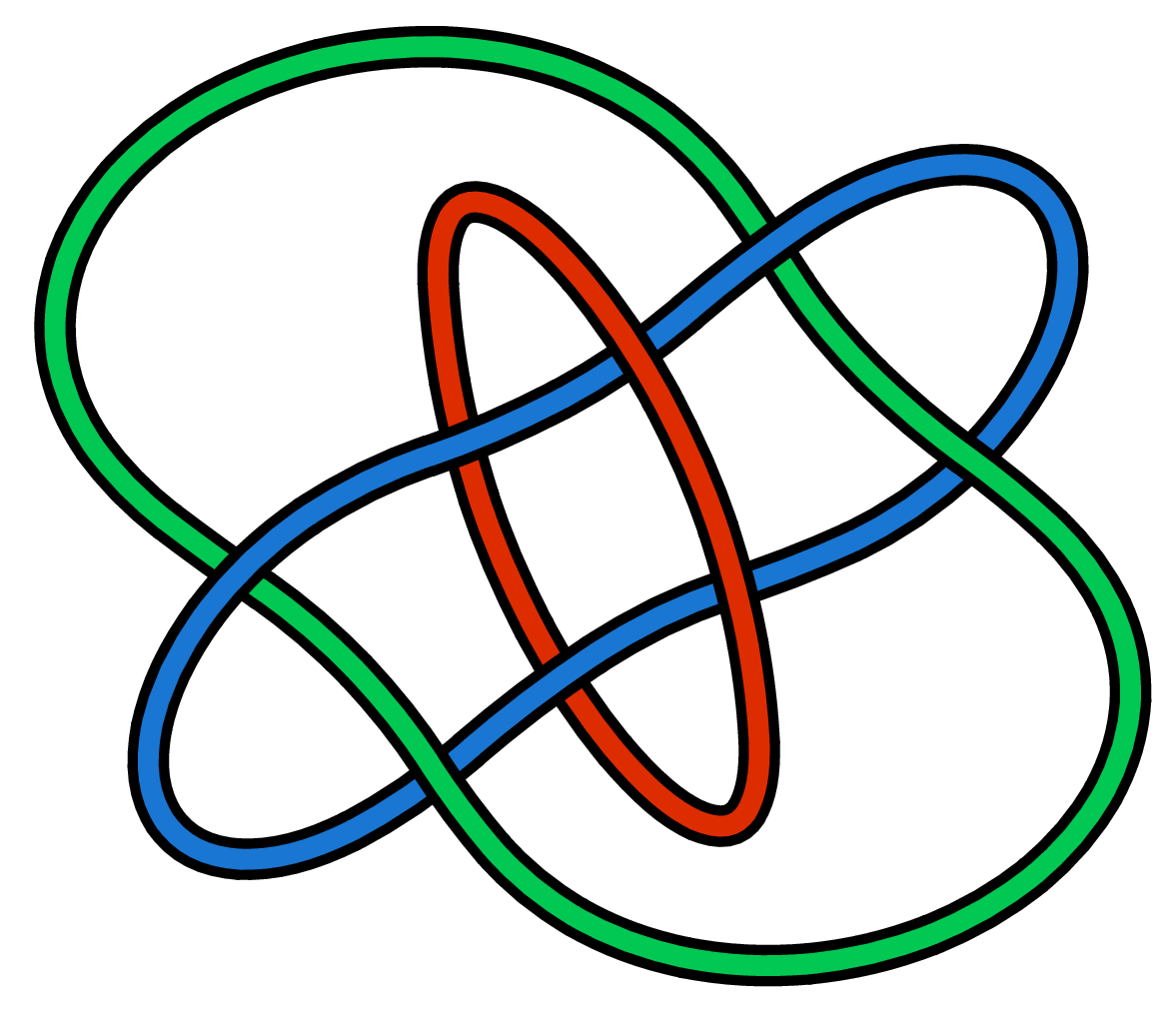} & \linkfig{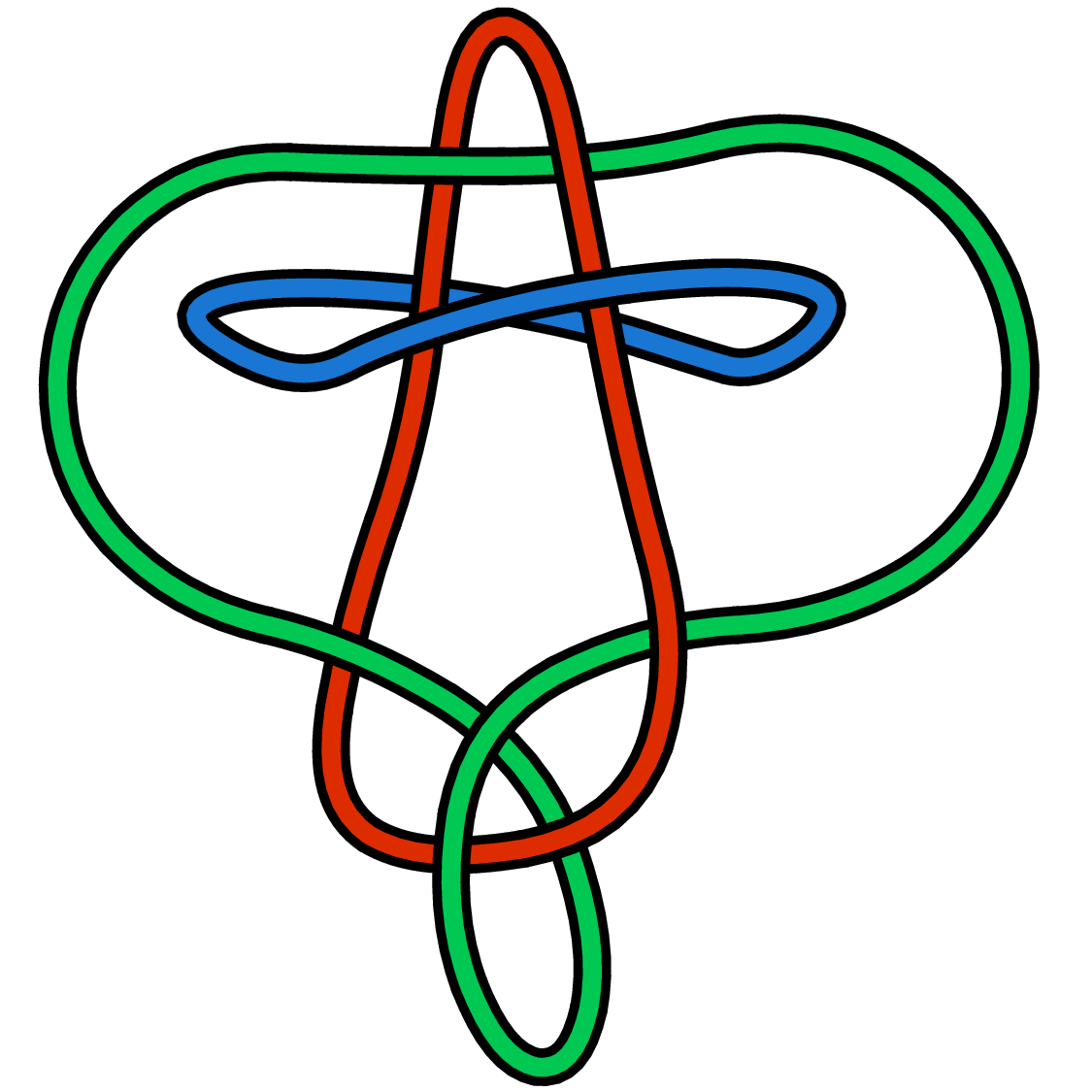} \\
(a) $3^1$: $\ra \rb \rc$ & (b) $3^2$: $\ra \rb \rc +\ra \rb$ \\[6pt]
\linkfig{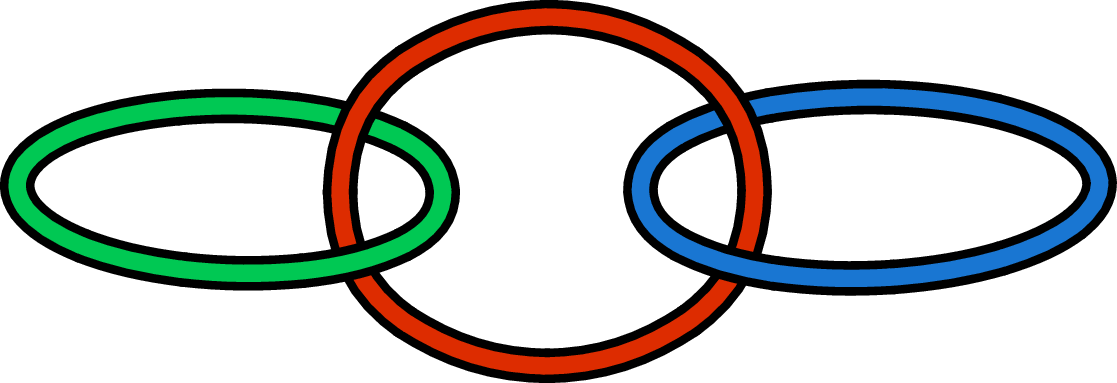} & \linkfig{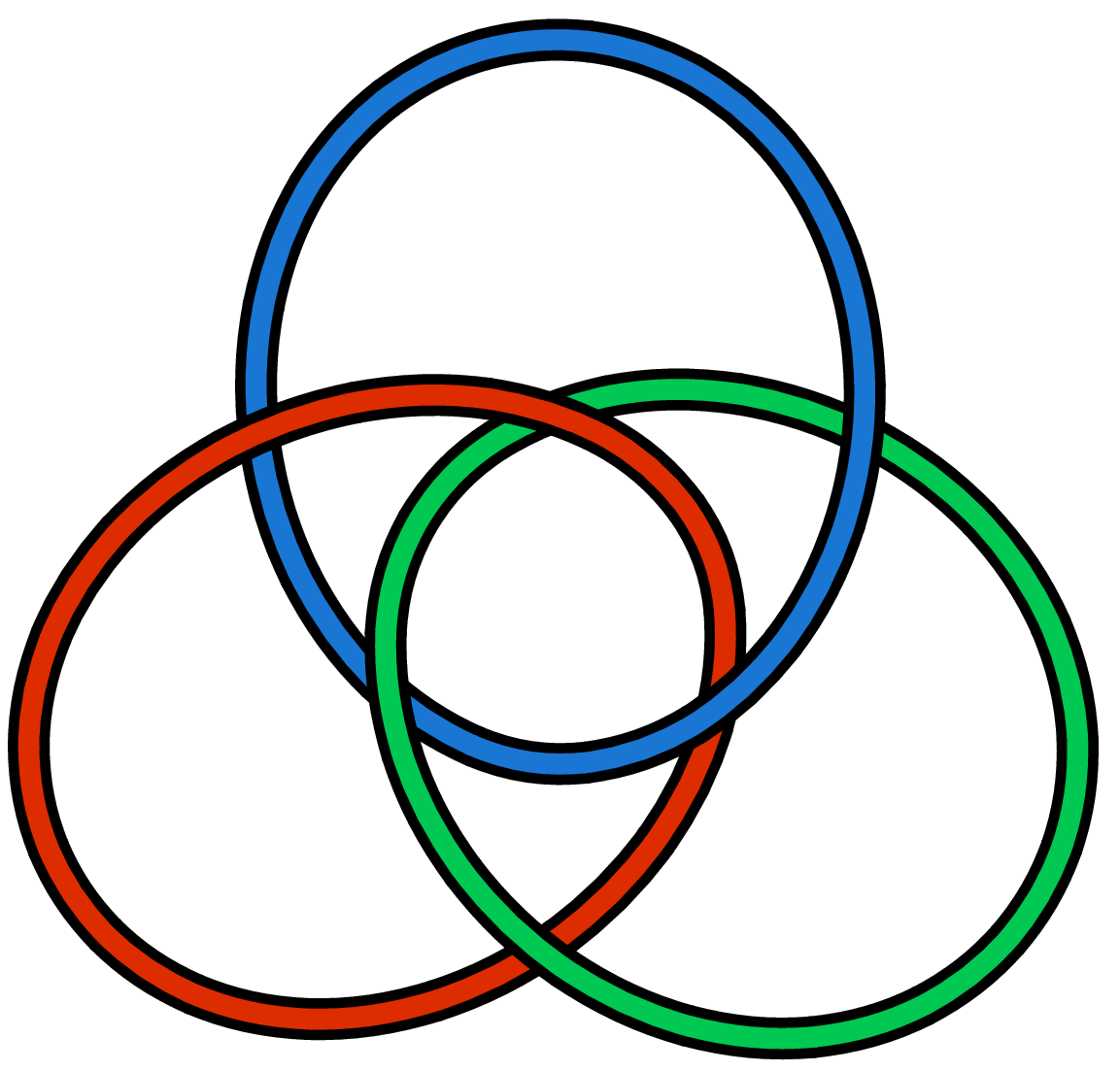} \\
(c) $3^3$: $\ra \rb +\ra \rc$ & (d) $3^4$: $\ra \rb +\ra \rc +\rb \rc$
\end{tabular}
\caption{Representatives of the four three-ring link classes and their polynomials. Rings $\ra$, $\rb$, and $\rc$ are drawn in red, green, and blue, respectively. The link class $3^1$ is commonly known as a Borromean link.}
\label{fig:three}
\end{figure}

The last rule removes information that is already present in the polynomial. For example, $ab+ac+abc$ reduces to $ab+ac$, because $ab$ and $ac$ already link $a$, $b$, and $c$, and every combination of cuts gives the same result for both polynomials. In contrast, $abcd$ cannot be dropped from $ab+cd+abcd$, since without it the link would split into two separate Hopf links. With these rules, each link class corresponds to exactly one polynomial, up to relabeling.

For three rings, these rules leave the four classes of Fig.~\ref{fig:three}, whose behavior under a single cut is summarized in Table~\ref{tab:cuts}. The Borromean link $3^1$ (in Fig.~\ref{fig:three}(a)) falls apart under any cut, while in $3^4$ every cut leaves a Hopf link. These two classes are represented by the GHZ and W states, respectively \cite{Dur2000,Quinta2018}. Pure three-qubit representatives of all four classes were given in \cite{vanDam2002, Quinta2018}. In the chain $3^3$, the rings stay linked only through the middle ring $a$. The class $3^2$ keeps $a$ and $b$ linked when $c$ is cut and falls apart under any other cut. For four rings there are 40 classes \cite{Quinta2018}.

\begin{table}[t]
\caption{Links remaining after cutting a single ring of each three-ring class; $0$ means no rings remain linked.}
\label{tab:cuts}
\begin{ruledtabular}
\begin{tabular}{lcccc}
Class & $\mathcal{P}$ & Cut $a$ & Cut $b$ & Cut $c$ \\ \hline
$3^1$ & $abc$ & $0$ & $0$ & $0$ \\
$3^2$ & $abc+ab$ & $0$ & $0$ & $ab$ \\
$3^3$ & $ab+ac$ & $0$ & $ac$ & $ab$ \\
$3^4$ & $ab+ac+bc$ & $bc$ & $ac$ & $ab$
\end{tabular}
\end{ruledtabular}
\end{table}

\section{From link polynomials to hypergraphs}\label{sec:hyper}

\begin{figure}[t]
\centering
\includegraphics[width=0.42\columnwidth]{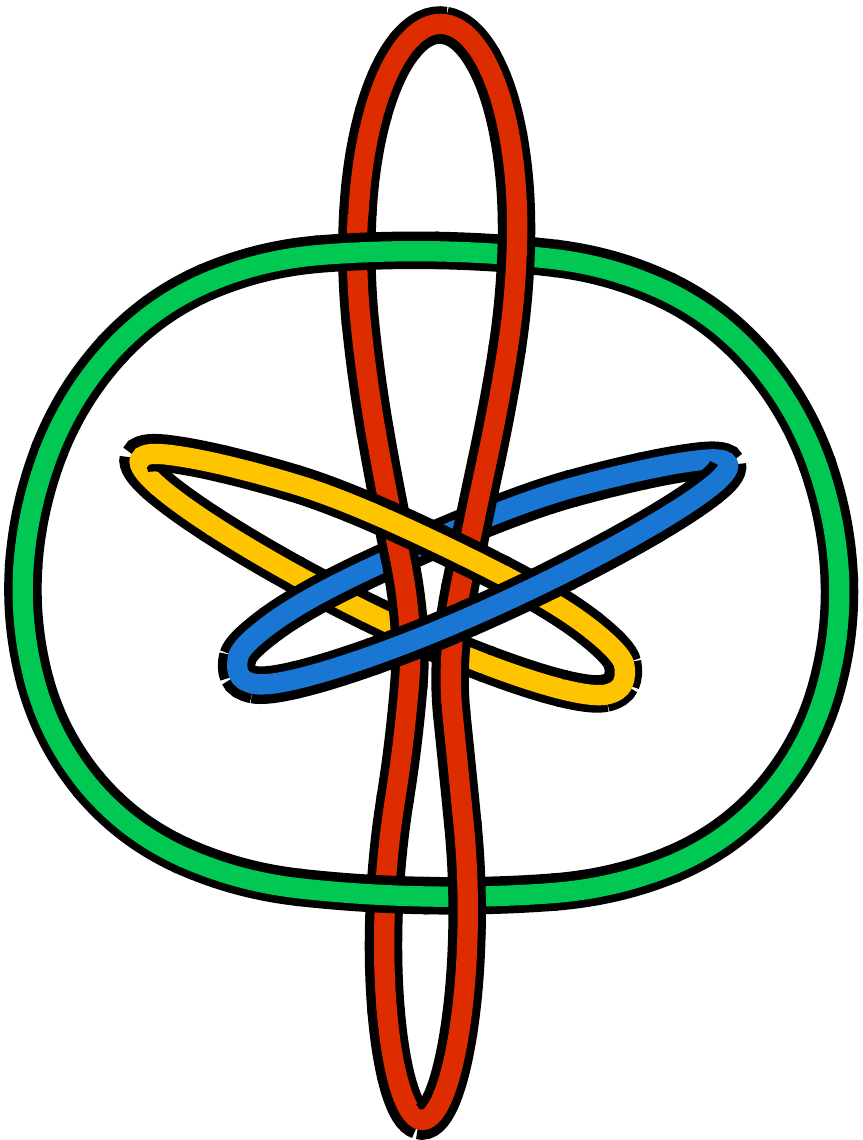}
\caption{An element of the class $4^{21}$, with polynomial $\ra \rb \rc +\ra \rb \rd +\rc \rd$. Rings $\ra$, $\rb$, $\rc$, and $\rd$ are drawn in red, green, blue, and yellow, respectively.}
\label{fig:421}
\end{figure}

The rules from the previous section can also be stated in the language of hypergraphs. Although this has already been noted in \cite{Honrubia2020}, we will review the mapping for completeness. A hypergraph $g=(V,E)$ consists of a set of vertices $V$ and a set $E$ of hyperedges, each hyperedge being a subset of $V$. Unlike an edge of an ordinary graph, a hyperedge can join any number of vertices. A polynomial defines a hypergraph whose vertices are the rings and whose hyperedges are the monomials, each monomial being identified with the set of rings it contains. By rule 3, every hyperedge has at least two vertices. We write $E_v=\{e\in E: v\in e\}$ for the hyperedges containing $v$, and $E_S=\{e\in E: e\subseteq S\}$ for those contained in $S\subseteq V$. As a running example, we use the four-ring class $4^{21}$ of Ref. \cite{Quinta2018}, shown in Fig.~\ref{fig:421}, with polynomial
\begin{equation}
\mathcal{P}(4^{21}) = abc+abd+cd\,.
\end{equation}
It has $V=\{a,b,c,d\}$ and, writing each hyperedge as its monomial, $E=\{abc,abd,cd\}$. We then have, for example, $E_a=E_b=\{abc,abd\}$, $E_c=\{abc,cd\}$, and $E_d=\{abd,cd\}$, while $E_{\{a,b,c\}}=\{abc\}$ and $E_{\{a,c,d\}}=\{cd\}$. In other words, $E_v$ is the set of hyperedges vertex $v$ connects to, while $E_S$ is the set of hyperedges that remains in the hypergraph after tracing out every vertex except for those in $S$.

Cutting the vertices outside a set $S$ deletes every hyperedge that touches them, leaving the induced hypergraph $g_S=(S,E_S)$. A bipartition $A|B$ of $S$ splits it into two nonempty disjoint sets $A$ and $B$ with $A\cup B=S$, and we say that a hyperedge $e$ crosses $A|B$ if it has vertices in both $A$ and $B$. The rings of $S$ remain one linked piece exactly when $g_S$ is connected, that is, when every bipartition of $S$ is crossed by some $e\in E_S$. In particular, by rule 2, the hypergraph $g=g_V$ of any link polynomial is connected. Each bipartition asks whether the rings in $A$ can be pulled away from those in $B$, which cannot happen if some hyperedge ties a ring of $A$ to a ring of $B$. In the example above, cutting ring $b$ leaves $S=\{a,c,d\}$ with $E_S=\{cd\}$. The set $S$ has three bipartitions, $a|cd$, $c|ad$, and $d|ac$. The hyperedge $cd$ crosses $c|ad$ and $d|ac$, since in each of them $c$ and $d$ lie on opposite sides, so these two splits are impossible. It does not cross $a|cd$, and nothing else ties $a$ to the pair $cd$, so ring $a$ slides off. Hence $c$ and $d$ remain linked while $a$ is freed, as expected from setting $b=0$ in the polynomial. A link class is therefore fully described by the family $\mathcal{K}(g)$ of subsets that stay linked or, in other words,
\begin{equation}\label{K}
\mathcal{K}(g)=\{S\subseteq V: S\neq\emptyset,\ g_S \text{ connected}\}.
\end{equation}
The polynomial rules have a direct meaning in terms of $\mathcal{K}$. A monomial is redundant precisely when the smaller hyperedges inside it already connect its vertices, and removing it leaves $\mathcal{K}$ unchanged. The reduced polynomial is therefore unique: its monomials are the members of $\mathcal{K}$ that cannot be obtained by joining smaller overlapping members. Two polynomials describe the same link class if and only if their families $\mathcal{K}$ coincide up to relabeling.

Families of this kind, containing all singletons and closed under unions of overlapping members, are known as connectivity spaces or connectedness systems \cite{Dugowson2010}. In the context of links and entanglement, they were introduced as proper correlation schemes in \cite{vanDam2002}. Every finite connectivity space can be realized by a link \cite{Debrunner1961, Kanenobu1986, Dugowson2010}, as also argued in \cite{vanDam2002, Quinta2019} through a braid construction, so every polynomial describes a genuine link class. A link class corresponds to a connected connectivity space, i.e., one that contains $V$ itself, and counting these up to relabeling gives $1, 4, 40, 3044$, and $26\,894\,586$ classes for $2$ to $6$ rings \cite{OEIS}.\footnote{The value for five rings corrects the value of 6900 quoted in Ref. \cite{Quinta2018}, which contains an inadvertent overcount of polynomials.} Including disconnected links, the counts become $2, 6, 47, 3095$, and $26\,897\,732$ \cite{vanDam2002, OEIS2}.

This viewpoint also clarifies what a quantum representative of a link class must do. Let $\rho_S$ be the reduced state on $S$, and call $S$ linked if $\rho_S$ is entangled across every bipartition of $S$. A state represents the class of $g$ when its linked subsets are exactly those in $\mathcal{K}(g)$. Since $\mathcal{K}(g)$ only depends on whether each $g_S$ is connected, it is enough to build, for any hypergraph, a state whose linked subsets are the connected ones (redundant hyperedges included). We now give such a construction with pure qudit states.

\section{From hypergraphs to pure qudit states}\label{sec:construction}

Let $g=(V,E)$ be a connected hypergraph. Each vertex $v$ is a qudit whose basis contains, for every $e\in E$, the labels $\ket{e^+}_v,\ket{e^-}_v$ if $v\in e$, or $\ket{e^0}_v$ if $v\notin e$. All labels are mutually orthogonal, so the dimension $d(v)$ of qudit $v$ equals the number of labels it carries,
\begin{equation}\label{dim}
d(v)=|E|+|E_v| .
\end{equation}
Since $|E_v|\leq|E|$, the local dimension satisfies $d(v)\leq2|E|$. The number of hyperedges can grow exponentially with $N=|V|$. It reaches $2^N-N-1$ only when redundant hyperedges are included, as in Corollary~\ref{cor:sat} below, but it can be exponential even for reduced polynomials; for the $m$-resistant classes of Sec.~\ref{sec:gme}, for instance, $|E|=\binom{N}{m}$, which scales as $2^N/\sqrt{N}$ for $m\approx N/2$. The local dimension is therefore exponential in $N$ in the worst case. Nevertheless, since $d(v)<2^{N+1}$, each qudit can be encoded in at most $N+1$ qubits, so the whole state requires at most $N(N+1)$ qubits. Writing $\ket{e^s}_X=\bigotimes_{v\in X}\ket{e^s}_v$, we define
\begin{equation}\label{g}
\ket{g}=\frac{1}{\sqrt{|E|}}\sum_{e\in E}\ket{\psi_e},\qquad
\ket{\psi_e}=\ket{e^0}_{V\setminus e}\otimes\ket{\mathrm{GHZ}_e},
\end{equation}
with $\ket{\mathrm{GHZ}_e}=(\ket{e^+}_e+\ket{e^-}_e)/\sqrt2$. Each hyperedge contributes a GHZ state on its vertices, while all other vertices hold the label $e^0$. These states should not be confused with the hypergraph states of Refs.~\cite{Rossi2013,Qu2013}, which are qubit states generated by multi-qubit controlled-$Z$ gates. The labels act as the classical flags commonly used in entanglement theory to make the components of a mixture locally distinguishable. To our knowledge, however, they have not been used before to realize prescribed patterns of entangled subsets, and the simplicity of the construction is precisely what allows the exact analysis below.

To see how the construction works in practice, we return to the class $4^{21}$ of Fig.~\ref{fig:421}, with polynomial $abc+abd+cd$, and build its state in four steps.

First, we read off the hyperedges from the monomials: $E=\{e_1,e_2,e_3\}$, with $e_1=abc$, $e_2=abd$, and $e_3=cd$.

Second, we assign the labels. Each qudit carries $e^\pm$ for the hyperedges it belongs to and $e^0$ for the others. For instance, qudit $d$ belongs to $e_2$ and $e_3$ but not to $e_1$, so its basis is $\{\ket{e_1^0},\ket{e_2^\pm},\ket{e_3^\pm}\}$. Every qudit belongs to exactly two of the three hyperedges, so Eq.~\eqref{dim} gives $d(v)=3+2=5$ for all of them.

Third, we write one term per hyperedge: a GHZ state on its rings, with every other ring holding the label $e^0$ of that hyperedge. Each ket has four slots, one per qudit, always in the order $a$, $b$, $c$, $d$, so that the corresponding qudits follow the order $\ket{x_a x_b x_c x_d}$. For example, in $\ket{e_2^+e_2^+e_2^0e_2^+}$ the qudits $a$, $b$, and $d$ hold $e_2^+$, while $c$, which is not in $e_2=abd$, holds $e_2^0$. The three terms are therefore
\begin{align}
\ket{\psi_{e_1}}&=\tfrac{1}{\sqrt2}\left(\ket{e_1^+e_1^+e_1^+e_1^0}+\ket{e_1^-e_1^-e_1^-e_1^0}\right),\nonumber\\
\ket{\psi_{e_2}}&=\tfrac{1}{\sqrt2}\left(\ket{e_2^+e_2^+e_2^0e_2^+}+\ket{e_2^-e_2^-e_2^0e_2^-}\right),\nonumber\\
\ket{\psi_{e_3}}&=\tfrac{1}{\sqrt2}\left(\ket{e_3^0e_3^0e_3^+e_3^+}+\ket{e_3^0e_3^0e_3^-e_3^-}\right).
\end{align}

Fourth, we superpose the three terms with equal weights, obtaining
\begin{equation}\label{Almost421}
\ket{4^{21}}=\frac{1}{\sqrt{3}}(\ket{\psi_{e_1}}+\ket{\psi_{e_2}}+\ket{\psi_{e_3}})\,.
\end{equation}
Finally, we replace the labels at each qudit by the integers $0,1,\dots,4$, numbered in order of appearance. For qudit $a$, for example, $e_1^+\to0$, $e_1^-\to1$, $e_2^+\to2$, $e_2^-\to3$, and $e_3^0\to4$. Applying this to Eq.~(\ref{Almost421}) immediately returns
\begin{align}\label{421num}
\ket{4^{21}}=\tfrac{1}{\sqrt6}\big(&\ket{0000}+\ket{1110}+\ket{2221}\nonumber\\
&+\ket{3322}+\ket{4433}+\ket{4444}\big).
\end{align}
It is straightforward to check that the state $\ket{4^{21}}$ reproduces the cuts of the corresponding link class. Tracing out qudit $b$ destroys the coherence of the GHZ states of $e_1$ and $e_2$ and leaves only the GHZ state of $e_3$ intact. Qudits $c$ and $d$ therefore remain entangled while $a$ becomes separable, exactly as found for the link in Sec.~\ref{sec:hyper}. Similarly, tracing out $c$ leaves only $e_2$ intact, so $a$, $b$, and $d$ remain entangled as in the Borromean link $abd$. The fact that this construction holds for any hypergraph leads to the following theorem:
\begin{theorem}\label{thm}
Let $g=(V,E)$ be a connected hypergraph. For every $S\subseteq V$ with $|S|\geq2$, $S$ is linked if and only if $g_S$ is connected. Moreover, if $g_S$ is not connected, $\rho_S$ contains no entanglement between the connected components of $g_S$: it is a mixture of product states with one factor per component.
\end{theorem}
The second statement means that any combination of cuts leaves precisely the direct sum of linked pieces predicted by the polynomial, in the sense of Ref.~\cite{Quinta2018}. For instance, for five rings with polynomial $abcde+ab+cd$, cutting $e$ leaves $ab+cd$, i.e., the link $2^1\oplus2^1$. Correspondingly, $\rho_{\{a,b,c,d\}}$ is separable across $ab|cd$, while each pair is entangled.
Connectivity of $g$ is needed only for $S=V$. For a disconnected hypergraph such as $ab+cd$, the pure state $\ket{g}$ is still entangled across the cut $ab|cd$, because the labels identifying the hyperedge are correlated coherently between the two components.
The proof, given in Appendix~\ref{app:proof}, rests on the observation that, in every term of Eq.~\eqref{g}, each qudit carries a label of the hyperedge that the term comes from, so each qudit locally records which hyperedge a term belongs to. As a consequence, tracing out a qudit turns every hyperedge containing it into classical noise, while the remaining hyperedges keep their GHZ entanglement and cannot interfere with one another. Separability is thus decided exactly, for any qudit dimension, by whether the hyperedges contained in $S$ cross a given bipartition. Redundant hyperedges do not change which subsets are linked, so every link polynomial, reduced or not, yields a pure state of its class with no coefficients to tune (they do, however, change the pattern of genuine multipartite entanglement, see Sec.~\ref{sec:gme}). Of course, this decoupling relies on each qudit being large enough to carry a separate set of labels for every hyperedge. With qubits there is no room for such labels, so different hyperedges inevitably interfere, and the problem becomes much harder: even for the special case of $m$-resistant states, the existence of qubit representatives for all $N$ and $m$ remains open~\cite{Quinta2019,Zhang2025,Han2026}. 

\section{A genuinely multipartite entangled construction}\label{sec:gme}

\begin{figure}[h!]
\centering
\includegraphics[width=\columnwidth]{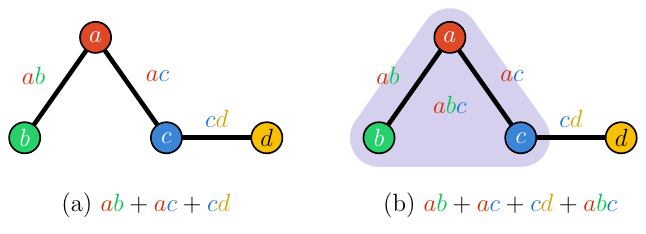}
\caption{Hypergraphs of (a) $\ra\rb+\ra\rc+\rc\rd$ and (b) $\ra\rb+\ra\rc+\rc\rd+\ra\rb\rc$. Both describe the same link class, since $\ra\rb\rc$ is redundant, but only in (b) is $\{\ra,\rb,\rc\}$ GME after cutting $\rd$. Two-ring monomials are drawn as lines and the three-ring monomial $\ra\rb\rc$ as a shaded region.}
\label{fig:gmehyper}
\end{figure}

Theorem~\ref{thm} characterizes when a subset $S$ is entangled across every bipartition. For the pure state $\ket{g}$ this is the strongest possible requirement, since a pure state entangled across every bipartition is genuinely multipartite entangled (GME). For the mixed reduced states $\rho_S$, however, it is weaker. A mixed state is biseparable if it can be written as a mixture of states, each of which is separable across some bipartition (possibly a different one for each state), and it is GME otherwise. A biseparable state can still be entangled across every bipartition, and this happens in our construction. For the star $ab+ac+ad$, the class $4^{35}$ of Ref.~\cite{Quinta2018}, cutting $d$ leaves $S=\{a,b,c\}$ linked through $ab$ and $ac$. However, $\rho_S$ is a mixture of a Bell pair on $ab$, a Bell pair on $ac$, and classical noise, each of which is separable across some bipartition, so $\rho_S$ is biseparable. The following theorem identifies exactly which subsets are GME.
\begin{theorem}\label{thm:gme}
Let $g=(V,E)$ be a connected hypergraph. For every $S\subsetneq V$ with $|S|\geq2$, $\rho_S$ is GME if and only if $S\in E$. Moreover, $\ket{g}$ is GME.
\end{theorem}
In other words, the state $\ket{g}$ encodes two patterns at once: the linked subsets are those connected by the hyperedges (Theorem~\ref{thm}), while the GME subsets are the hyperedges themselves, together with $V$. This gives rule 5 a quantum meaning. Polynomials that differ only by redundant monomials describe the same link, but their states have different GME patterns. For instance, the hypergraphs $ab+ac+cd$ and $ab+ac+cd+abc$ of Fig.~\ref{fig:gmehyper} describe the same four-ring link, but after cutting $d$ only the second case returns a GME state. To see this explicitly, we list the hyperedges of the second case as $e_1=ab$, $e_2=ac$, $e_3=cd$, and $e_4=abc$, and number the labels at each qudit in order of appearance, as in Sec.~\ref{sec:construction}. The local dimensions are then $(7,6,7,5)$ for qudits $(a,b,c,d)$, and Eq.~\eqref{g} gives
\begin{align}\label{gmestate}
\ket{g}=\tfrac{1}{\sqrt8}\big(&\ket{0000}+\ket{1100}+\ket{2211}+\ket{3221}\nonumber\\
&+\ket{4332}+\ket{4343}+\ket{5454}+\ket{6564}\big),
\end{align}
where consecutive pairs of terms correspond to $e_1$, $e_2$, $e_3$, and $e_4$, respectively. Since the labels of different hyperedges are orthogonal at qudit $d$, tracing it out removes all interference between hyperedges and leaves
\begin{align}\label{gmerho}
\rho_{abc}=\;&\tfrac14\rho(\ket{\phi_1})+\tfrac14\rho(\ket{\phi_2})+\tfrac18\rho(\ket{433})\nonumber\\
&+\tfrac18\rho(\ket{434})+\tfrac14\rho(\ket{\phi_3}),
\end{align}
with $\rho(\ket{\phi})=\ket{\phi}\bra{\phi}$, $\ket{\phi_1}=(\ket{000}+\ket{110})/\sqrt2$, $\ket{\phi_2}=(\ket{221}+\ket{322})/\sqrt2$, and $\ket{\phi_3}=(\ket{545}+\ket{656})/\sqrt2$. The states $\ket{\phi_1}$ and $\ket{\phi_2}$ are Bell pairs on $ab$ and $ac$, in a product state with $c$ and $b$, respectively, and $\ket{433}$ and $\ket{434}$ are product states. The component $\ket{\phi_3}$, instead, is a GHZ state on $a$, $b$, and $c$, contributed by $e_4=abc$, and it makes $\rho_{abc}$ GME. For $ab+ac+cd$, the state consists of the first six terms of Eq.~\eqref{gmestate} only, with normalization $1/\sqrt6$, and the corresponding $\rho_{abc}$ lacks the component $\ket{\phi_3}$. It is then a mixture of states that are each separable across some bipartition, and hence biseparable.

The reduced polynomial therefore yields the leanest state, which is GME only on its monomials. At the other extreme, we may include every linked subset as a hyperedge. Let $\bar{E}=\{S\in\mathcal{K}(g): |S|\geq2\}$ be the family of all linked subsets and $\bar g=(V,\bar E)$ the corresponding hypergraph.
\begin{corollary}\label{cor:sat}
The hypergraph $\bar g$ represents the same link class as $g$, and in the state $\ket{\bar g}$ every subset $S$ with $|S|\geq2$ is GME if and only if it is linked.
\end{corollary}
Every link class thus admits a pure state in which each group of particles that stays linked after any pattern of losses is GME, while every other group falls apart. The price is a larger local dimension, $\bar d(v)=|\bar E|+|\bar E_v|$. For the star, $\bar E$ adds $abc$, $abd$, $acd$, and $abcd$ to the three original monomials, raising the dimensions from $(6,4,4,4)$ to $(14,11,11,11)$ for qudits $(a,b,c,d)$. For the running example $4^{21}$, no change is needed. Its only linked subset that is not a monomial is $V$ itself, and the pure state is GME in any case, so the state of Eq.~\eqref{421num} is already GME on every linked subset.

Theorem~\ref{thm:gme} also settles the existence of the $m$-resistant states discussed in the Introduction. For $0\leq m\leq N-2$, the $m$-resistant link of $N$ rings has as polynomial the sum of all $\binom{N}{N-m}$ monomials with $N-m$ variables, i.e., the hypergraph whose hyperedges are all subsets of size $N-m$.
\begin{corollary}\label{cor:mres}
For every $N$ and $0\leq m\leq N-2$, let $g$ be the hypergraph whose hyperedges are all subsets of $V$ of size $N-m$. After the loss of any $m$ particles, the remaining particles of $\ket{g}$ are GME, while after the loss of any $m+1$ particles they are fully separable. The local dimension is $d(v)=\binom{N}{m}+\binom{N-1}{m}$.
\end{corollary}
This proves the conjecture of Ref.~\cite{Quinta2019} that $m$-resistant pure states exist for all $N$ and $m$ in some local dimension, previously established for $m\leq(N-2)/2$, i.e., $N\geq2(m+1)$~\cite{Quinta2019}, and in some qubit cases beyond this range~\cite{Zhang2025}. It does so in the strong sense of Ref.~\cite{Zhang2025}, in which the remaining particles are GME. Note that after the loss of $k$ particles with $0<k<m$, the remaining set is linked but is not a hyperedge, so by Theorem~\ref{thm:gme} it is only biseparable. If GME is required after any $k\leq m$ losses, Corollary~\ref{cor:sat} applies: taking every subset of size between $N-m$ and $N-1$ as a hyperedge preserves the link class, makes the remaining particles GME after any $k\leq m$ losses, and leaves them fully separable after any $m+1$, at the cost of the local dimension $d(v)=\sum_{j=1}^{m}\big[\binom{N}{j}+\binom{N-1}{j}\big]$. Whether qubits suffice for all $N$ and $m$ remains open~\cite{Zhang2025,Han2026}.

Theorem~\ref{thm:gme} also suggests a refinement of the link classification, in which states are distinguished by both their linked and their GME subsets. By Theorems~\ref{thm} and~\ref{thm:gme}, the GME subsets $S\subsetneq V$ of $\ket{g}$ are exactly the hyperedges in $F=E\setminus\{V\}$, and these determine the linked subsets as well. Two states $\ket{g}$ and $\ket{g'}$ therefore share both patterns if and only if $F$ and $F'$ coincide up to relabeling. The full set $V$ plays no role, since it is always linked and GME, and adding it as a hyperedge makes any hypergraph connected without changing either pattern. The refined classes are thus obtained by dropping rule 5 and identifying polynomials that differ only by the monomial containing all rings, so that they are labeled by an arbitrary family $F$ of subsets with sizes between $2$ and $N-1$. Unlike link classes, they can be counted in closed form. There are $2^{\,2^N-N-2}$ labeled classes, while, by Burnside's lemma, the number up to relabeling is
\begin{equation}\label{gmecount}
\frac{1}{N!}\sum_{\pi\in S_N}2^{\,c(\pi)},
\end{equation}
where $c(\pi)$ is the number of cycles of the permutation $\pi$ acting on the subsets of $V$ of sizes $2$ to $N-1$. Equation~\eqref{gmecount} gives $1$, $4$, $90$ and $306\,016$ classes for $2$ to $5$ rings. For three rings nothing changes, because the only possible redundant monomial is $abc$ itself. For four rings, each of the $40$ link classes splits into one or more refined classes, $90$ in total. The chain of Fig.~\ref{fig:gmehyper}, for instance, has two proper linked subsets that are not monomials, $abc$ and $acd$, and splits into three classes: no GME triple, one GME triple ($abc$ or, equivalently by the symmetry of the chain, $acd$), or both. Note that dropping rule 5 overcounts link classes, so Eq.~\eqref{gmecount} counts the finer patterns rather than links. The reduced polynomial and the saturated hypergraph $\bar g$ of Corollary~\ref{cor:sat} are the two extremes within each link class, and the latter provides a canonical choice of one GME-saturated state per link class.

\section{Conclusions}\label{sec:conclusions}

We have shown that every link class, and hence every pattern of linked subsets  allowed by the link formalism, is realized by a pure qudit state. The construction is read directly off the link polynomial: each monomial becomes a GHZ state on its rings, every ring carries a qudit label identifying the monomial and the state is the equal superposition of all monomials. Tracing out a qudit then destroys exactly the monomials containing it, so a subset of qudits stays entangled across all its bipartitions if and only if the surviving monomials still link it. Along the way, recasting link polynomials as hypergraphs identifies link classes with connectivity spaces, whose known enumeration gives $1,4,40,3044$ and $26\,894\,586$ classes for two to six rings.

The same construction also controls genuine multipartite entanglement. A subset that survives the loss of the others is GME exactly when it is itself a monomial, so polynomials describing the same link can yield states with different GME patterns. Including every linked subset as a monomial gives states in which every linked group of particles is GME. In particular, $m$-resistant pure states, whose remaining particles are GME after any $m$ losses and fully separable after any $m+1$, exist for every $N$ and $m$.

The main open question is efficiency: the local dimension can grow exponentially with $N$, corresponding to up to $N+1$ qubits per party, while qubit representatives are known only for specific classes~\cite{Quinta2018,Zhang2025,Han2026}. Whether qubits suffice for every link class, or even for every $m$-resistant one, remains open. Finding the minimal local dimension for each class and states that are easy to prepare in the laboratory, are natural next steps toward quantum networks with a prescribed pattern of shared entanglement.

\begin{acknowledgments}
We thank Wim van Dam for helpful correspondence and for bringing his earlier work to our attention.
\end{acknowledgments}

\appendix

\section{Proofs}\label{app:proof}

\subsection{Proof of Theorem~\ref{thm}}

\begin{proof}[Proof of Theorem~\ref{thm}]
We use throughout that the labels of different hyperedges are orthogonal at every qudit. Consider first $S\neq V$. Expanding $\ket{g}\bra{g}$, every term $\ket{\psi_e}\bra{\psi_{e'}}$ with $e\neq e'$ vanishes after trace, since at any traced vertex the labels of $e$ and $e'$ are orthogonal, thus
\begin{equation}\label{rhoS}
\rho_S=\frac{1}{|E|}\sum_{e\in E}\sigma_e,\qquad \sigma_e=\mathrm{Tr}_{V\setminus S}\ket{\psi_e}\bra{\psi_e}.
\end{equation}
Each $\sigma_e$ is supported only on the labels of $e$, at every vertex of $S$. Consequently, both $\rho_S$ and its partial transpose $\rho_S\PT{A}$ are block diagonal, with one block per hyperedge. Because the blocks carry locally orthogonal labels, $\rho_S$ is separable across $A|B$ iff every block $\sigma_e$ is and $\rho_S\PT{A}$ has a negative eigenvalue iff some $\sigma_e\PT{A}$ does. (For the ``only if'' part of the first statement, projecting every qudit locally onto the labels of $e$ maps $\rho_S$ to $\sigma_e/|E|$ and preserves separability.)

It remains to examine a single hyperedge. If $e\not\subseteq S$, some vertex of $e$ is traced out, which destroys the GHZ coherence:
\begin{equation}\label{A2}
\sigma_e=\frac12\sum_{s=\pm}\rho\!\left(\ket{e^s}_{e\cap S}\ket{e^0}_{S\setminus e}\right),
\end{equation}
where $\rho(\ket{\phi})=\ket{\phi}\bra{\phi}$. This is a mixture of product states: cutting a ring removes every hyperedge it belongs to. If instead $e\subseteq S$, then $\sigma_e=\rho(\ket{\mathrm{GHZ}_e}\ket{e^0}_{S\setminus e})$. When $e$ lies within $A$ or within $B$, this is a product state across $A|B$. When $e$ crosses $A|B$, we write $\ket{\pm}_A=\ket{e^\pm}_{e\cap A}$ and $\ket{\pm}_B=\ket{e^\pm}_{e\cap B}$, so that $\ket{\mathrm{GHZ}_e}=(\ket{++}+\ket{--})/\sqrt2$ is a Bell state between the two sides. Omitting the labels $e^0$, which play no role, the partial transpose on $A$ is
\begin{align}
\sigma_e\PT{A}&=\tfrac12\big(\ket{++}\bra{++}+\ket{--}\bra{--}\nonumber\\
&\qquad+\ket{-+}\bra{+-}+\ket{+-}\bra{-+}\big)\nonumber\\
&=\tfrac12\rho(\ket{++})+\tfrac12\rho(\ket{--})\nonumber\\
&\quad+\tfrac12\rho(\ket{\Psi^+})-\tfrac12\rho(\ket{\Psi^-}),\label{PT}
\end{align}
with $\ket{\Psi^\pm}=(\ket{+-}\pm\ket{-+})/\sqrt2$. Since Eq.~\eqref{PT} exhibits the negative eigenvalue $-\tfrac12$, with eigenvector $\ket{\Psi^-}$, we can combine these results for $S\neq V$ to obtain the criterion
\begin{equation}\label{crit}
\rho_S \text{ entangled across } A|B \iff E_S \text{ crosses } A|B ,
\end{equation}
where ``$E_S$ crosses $A|B$'' means that some $e\in E_S$ crosses $A|B$. If such an $e$ exists, $\rho_S\PT{A}$ has the eigenvalue $-1/(2|E|)$, so $\rho_S$ is entangled \cite{Peres1996}. Otherwise every block is separable across $A|B$ and so is $\rho_S$.

By definition, $S$ is linked if $\rho_S$ is entangled across every bipartition of $S$ and $g_S$ is connected if every bipartition of $S$ is crossed by some $e\in E_S$. Equation~\eqref{crit} shows that these two conditions agree for each bipartition separately, so they also agree for all bipartitions at once: $S$ is linked iff $g_S$ is connected.

To prove the second statement of the theorem for $S\neq V$, let $C_1,\dots,C_k$ be the connected components of $g_S$, including isolated vertices. A hyperedge $e\subseteq S$ is connected by itself, so it lies inside a single component and its block $\sigma_e=\rho(\ket{\mathrm{GHZ}_e})\otimes\rho(\ket{e^0}_{S\setminus e})$ is a product across $C_1|\cdots|C_k$. A hyperedge $e\not\subseteq S$ gives the block of Eq.~\eqref{A2}, a mixture of product states over all qudits of $S$. Every block is thus a mixture of product states with one factor per component and so is their mixture $\rho_S$, which therefore contains no entanglement between different components. For $S=V$, the hypergraph $g$ is connected and the statement is empty.

Finally, for $S=V$, the only case where connectivity of $g$ is used, the state is pure and it is entangled across $A|B$ iff $\rho_A$ is mixed. Since $B\neq\emptyset$, Eq.~\eqref{rhoS} applies to $\rho_A$, which therefore has $|E|$ orthogonal blocks of trace $1/|E|$ each and is mixed whenever $|E|\geq2$. If $|E|=1$, connectivity forces the single hyperedge to be $V$ itself and $\ket{g}=\ket{\mathrm{GHZ}_V}$ is entangled across every bipartition. As $g$ is connected, $V$ is both connected and linked.
\end{proof}

\subsection{Proof of Theorem~\ref{thm:gme} and Corollaries~\ref{cor:sat} and~\ref{cor:mres}}

A state $\rho$ of the qudits in $S$ is biseparable if $\rho=\sum_k p_k\rho_k$ with $p_k\geq0$, where each $\rho_k$ is separable across some bipartition $A_k|B_k$ of $S$. It is GME otherwise.

\begin{proof}[Proof of Theorem~\ref{thm:gme}]
Let $S\subsetneq V$ with $|S|\geq2$, so that Eq.~\eqref{rhoS} holds. First we show that if $S\notin E$, then $\rho_S$ is biseparable. Consider each block $\sigma_e$ in Eq.~\eqref{rhoS}. If $e\not\subseteq S$, Eq.~\eqref{A2} shows that $\sigma_e$ is a mixture of product states over all qudits of $S$, hence separable across every bipartition. If $e\subseteq S$, then $e\neq S$ because $S\notin E$, so both $e$ and $S\setminus e$ are nonempty and
\begin{equation}
\sigma_e=\rho\big(\ket{\mathrm{GHZ}_e}\big)\otimes\rho\big(\ket{e^0}_{S\setminus e}\big)
\end{equation}
is a product state across the bipartition $e\,|\,S\setminus e$. Every block is therefore separable across some bipartition and $\rho_S$, being a mixture of such blocks, is biseparable. Instead, if $S\in E$, then $\rho_S$ is GME. For each $v\in S$, let
\begin{equation}
\Pi_v=\ket{S^+}\bra{S^+}_v+\ket{S^-}\bra{S^-}_v
\end{equation}
be the projector onto the labels of the hyperedge $S$ at qudit $v$ and let $F=\bigotimes_{v\in S}\Pi_v$. For $e\neq S$, the block $\sigma_e$ is supported on the labels of $e$ at every qudit of $S$, which are orthogonal to those of $S$, so $F\sigma_eF=0$. For $e=S$, we have $\sigma_S=\rho(\ket{\mathrm{GHZ}_S})$ and $F\sigma_SF=\sigma_S$. Hence
\begin{equation}\label{filter}
F\rho_SF=\frac{1}{|E|}\,\rho\big(\ket{\mathrm{GHZ}_S}\big).
\end{equation}
Suppose now that $\rho_S=\sum_kp_k\rho_k$ were biseparable. Since $F$ is a tensor product of local operators, each $F\rho_kF$ is still separable across $A_k|B_k$. Equation~\eqref{filter} then writes the pure state $\rho(\ket{\mathrm{GHZ}_S})$ as a sum of positive operators $|E|\,p_kF\rho_kF$. Because a pure state cannot be decomposed into distinct states, every nonzero term must be proportional to $\rho(\ket{\mathrm{GHZ}_S})$, which would then be separable across some $A_k|B_k$. This is impossible, since across any bipartition $A|B$ of $S$ we have $\ket{\mathrm{GHZ}_S}=(\ket{++}+\ket{--})/\sqrt2$ in the notation of Eq.~\eqref{PT}, which has Schmidt rank two. Therefore $\rho_S$ is GME. Finally, $\ket{g}$ is pure and, by Theorem~\ref{thm}, entangled across every bipartition of $V$, hence GME.
\end{proof}

\begin{proof}[Proof of Corollary~\ref{cor:sat}]
We first show that $\mathcal{K}(\bar g)=\mathcal{K}(g)$. Every hyperedge of $g$ is connected in $g$, so $E\subseteq\bar E$ and every $S$ connected in $g$ is connected in $\bar g$. Conversely, let $S$ be connected in $\bar g$. Then the hyperedges of $\bar g$ inside $S$ can be added one at a time, each overlapping the union of the previous ones, until they cover $S$. Each of them is connected in $g$ through hyperedges of $g$ lying inside it, hence inside $S$. Since the union of two overlapping sets that are connected in $g_S$ is again connected in $g_S$, the whole of $S$ is connected in $g$. Thus $g$ and $\bar g$ represent the same link class. By Theorem~\ref{thm:gme} applied to $\bar g$, for $S\neq V$ the state $\rho_S$ is GME iff $S\in\bar E$, that is, iff $S$ is linked. For $S=V$, $\ket{\bar g}$ is GME and $V$ is linked.
\end{proof}

\begin{proof}[Proof of Corollary~\ref{cor:mres}]
After the loss of $m$ particles, the remaining set $S$ has $N-m$ elements and is itself a hyperedge, so $\rho_S$ is GME by Theorem~\ref{thm:gme} (for $m=0$, $\ket{g}=\ket{\mathrm{GHZ}_V}$). After the loss of $m+1$ particles, $S$ has $N-m-1$ elements and contains no hyperedge, so every block in Eq.~\eqref{rhoS} is of the form~\eqref{A2}, a mixture of product states over all qudits of $S$. Hence $\rho_S$ is fully separable. The local dimension follows from Eq.~\eqref{dim} with $|E|=\binom{N}{N-m}=\binom{N}{m}$ and $|E_v|=\binom{N-1}{N-m-1}=\binom{N-1}{m}$.
\end{proof}


\begin{thebibliography}{99}

\bibitem{Dur2000}
W. D\"ur, G. Vidal and J.~I. Cirac, Phys. Rev. A \textbf{62}, 062314 (2000).

\bibitem{Verstraete2002}
F. Verstraete, J. Dehaene, B. De Moor, and H. Verschelde, Phys. Rev. A \textbf{65}, 052112 (2002).

\bibitem{Seveso2018}
L. Seveso, D. Goyeneche, and K. \.{Z}yczkowski, J. Math. Phys. \textbf{59}, 072203 (2018).

\bibitem{Burchardt2021}
A. Burchardt, Ph.D. thesis, Jagiellonian University, 2021; arXiv:2204.13441.

\bibitem{Melnikov2023}
D. Melnikov, Phys. Rev. D \textbf{107}, 126005 (2023).

\bibitem{Mengoni2020}
R. Mengoni, A. di Pierro, L. Memarzadeh, and S. Mancini, Quantum Inf. Comput. \textbf{20}, 375 (2020).

\bibitem{Quinta2018}
G.~M. Quinta and R. Andr\'e, Phys. Rev. A \textbf{97}, 042307 (2018).

\bibitem{Aravind1997}
P.~K. Aravind, in \emph{Potentiality, Entanglement and Passion-at-a-Distance}, edited by R.~S. Cohen, M. Horne, and J. Stachel (Kluwer, Dordrecht, 1997), p.~53.

\bibitem{vanDam2002}
W. van Dam, ``Qualifying entanglement with knot theory'', talk at the Workshop on Quantum Information Processing (QIP 2003), MSRI, Berkeley, December (2002); \url{legacy.slmath.org/workshops/204/schedules/1241}

\bibitem{Sugita2006}
A. Sugita, in \emph{Proceedings of the International Workshop on Knot Theory for Scientific Objects} (Osaka, Japan, 2006).

\bibitem{Kauffman2002}
L. H. Kauffman and S. J. Lomonaco, New J. Phys. \textbf{4}, 73 (2002).

\bibitem{Padmanabhan2020a}
P. Padmanabhan, F. Sugino, and D. Trancanelli, Quantum Inf. Comput. \textbf{20}, 37 (2020).

\bibitem{Padmanabhan2020b}
P. Padmanabhan, F. Sugino, and D. Trancanelli, Quantum \textbf{4}, 311 (2020).

\bibitem{Padmanabhan2020c}
P. Padmanabhan, F. Sugino, and D. Trancanelli, Quantum Inf. Comput. \textbf{20}, 1154 (2020).

\bibitem{Padmanabhan2021}
P. Padmanabhan, F. Sugino, and D. Trancanelli, J. Phys. A \textbf{54}, 135301 (2021).

\bibitem{Zhang2024}
K. Zhang, K. Hao, K. Yu, V. Korepin, and W.-L. Yang, J. Phys. A \textbf{57}, 445303 (2024).

\bibitem{Padmanabhan2026}
P. Padmanabhan, S. Maity, A. Sinha, and V. Korepin, Open Commun. Nonlinear Math. Phys. (2026); arXiv:2606.26510.

\bibitem{Bhattacharyya2026}
S. Bhattacharyya and S. Roy, arXiv:2608.09076.

\bibitem{Quinta2019}
G. M. Quinta, R. Andr\'e, A. Burchardt, and K. \.{Z}yczkowski, Phys. Rev. A \textbf{100}, 062329 (2019).

\bibitem{Zhang2025}
W. Zhang, Z. Han, F. Shi, and X. Zhang, Phys. Rev. A \textbf{112}, 022408 (2025).

\bibitem{Han2026}
Z. Han, W. Zhang, and X. Zhang, arXiv:2606.08561.

\bibitem{Plesch2003}
M. Plesch and V. Bu\v{z}ek, Phys. Rev. A \textbf{67}, 012322 (2003).

\bibitem{Gharahi2016}
M. Gharahi Ghahi and S. J. Akhtarshenas, Eur. Phys. J. D \textbf{70}, 54 (2016).

\bibitem{Honrubia2020}
E. Honrubia and \'{A}. S. Sanz, Quantum Rep. \textbf{2}, 352 (2020).

\bibitem{Dugowson2010}
S. Dugowson, Cah. Topol. G\'eom. Diff\'er. Cat\'eg. \textbf{51}, 282 (2010); arXiv:1001.2378.

\bibitem{Debrunner1961}
H. Debrunner, Duke Math. J. \textbf{28}, 17 (1961).

\bibitem{Kanenobu1986}
T. Kanenobu, J. Math. Soc. Japan \textbf{38}, 295 (1986).

\bibitem{OEIS}
OEIS Foundation Inc., The On-Line Encyclopedia of Integer Sequences, Sequence A072445, \url{https://oeis.org/A072445}.

\bibitem{OEIS2}
OEIS Foundation Inc., The On-Line Encyclopedia of Integer Sequences, Sequence A072444, \url{https://oeis.org/A072444}

\bibitem{Rossi2013}
M. Rossi, M. Huber, D. Bru\ss, and C. Macchiavello, New J. Phys. \textbf{15}, 113022 (2013).

\bibitem{Qu2013}
R. Qu, J. Wang, Z.-S. Li, and Y.-R. Bao, Phys. Rev. A \textbf{87}, 022311 (2013).

\bibitem{Peres1996}
A. Peres, Phys. Rev. Lett. \textbf{77}, 1413 (1996).

\end{thebibliography}
\end{document}